\documentclass[12pt,preprint]{aastex}

\usepackage{natbib}
\usepackage{epsfig}

\newcommand{\rhill}{r_{\mathrm{H}}}
\newcommand{\msun}{\:\mathrm{M}_{\sun}}
\newcommand{\mearth}{\ensuremath{\mathrm{M}_{\earth}}}

\newcommand{\figref}[1]{Figure \ref{#1}}
\newcommand{\eqref}[1]{equation (\ref{#1})}

\newcommand{\paperone}{Paper I\@}

\shortauthors{Jang-Condell \& Sasselov}

\begin{document}
\slugcomment{To appear in the Astrophysical Journal}
\title{Disk Temperature Variations and Effects on the Snow Line 
in the Presence of Small Protoplanets
}
\author{Hannah Jang-Condell\altaffilmark{1}
\and Dimitar D.~Sasselov\altaffilmark{2}}
\affil{Harvard-Smithsonian Center for Astrophysics}
\affil{60 Garden St., Cambridge, MA 02138}
\altaffiltext{1}{hjang@cfa.harvard.edu}
\altaffiltext{2}{dsasselov@cfa.harvard.edu}

\begin{abstract}
We revisit the computation of a ``snow line'' in a passive protoplanetary 
disk during the stage of planetesimal formation.  
We examine how shadowing and illumination in the vicinity 
of a planet affects where in the disk ice can form, making 
use of our method for calculating radiative transfer on disk perturbations 
with some improvements on the model.  We adopt a model for the 
unperturbed disk structure that is more consistent with 
observations and use opacities for reprocessed dust instead of 
interstellar medium 
dust.  We use the improved disk model to calculate the temperature 
variation for a range of planet masses and distances and find that 
planets at the gap-opening threshold can induce temperature variations 
of up to $\pm30\%$.  Temperature variations this significant may 
have ramifications for planetary accretion rates and migration rates.  
We discuss in particular the effect of temperature variations 
near the sublimation point of water, since the formation of ice 
can enhance the accretion rate of disk material onto a planet.  
Shadowing effects can cool the disk enough that ice will form 
closer to the star than previously expected, effectively moving the 
snow line inward.  
\end{abstract}

\section{Introduction}

The concept of a ``snow line'' was introduced by \citet{hayashi} 
and refers to the distance from the Sun at which the midplane 
temperature of the preplanetary solar nebula drops to the sublimation 
temperature of ice.  The presence of ice beyond the snow line, which 
Hayashi calculated to be at 2.7 AU, ought to enhance planet formation 
and explains the existence of the gas and ice giants 
in their present locations.  \citet{snowline} revisited this 
issue, adopting an updated protoplanetary disk model in hydrostatic 
and radiative equilibrium, with little or no accretion heating 
(passive disk).  
They find that 
the snow line can be as close as $\sim1$ AU, depending on 
disk parameters.  However, if the temperature of the disk varies 
with height \citep[e.g.,][]{vertstruct}, different parts of the disk 
will reach the sublimation temperature of water, 170 K, 
at different radii, so rather than a snow ``line'' 
in the disk, there will be a snow ``transition.''  In this paper, 
we show that the presence of planets themselves can affect 
where in the disk the snow transition occurs.  

We have previously shown that the presence of a protoplanet 
can affect the temperature structure in the protoplanetary disk 
\citep[hereafter \paperone{}]{paper1}.  These temperature variations 
affect where in the disk ice can form.  The goal of this study is 
to quantify this effect by undertaking a parameter 
study in which we calculate the temperature structure for a variety 
of planet masses and distances. 

This work is a step toward bridging the gap between simulations of 
disk-protoplanet interactions and analytical models based on 
observations of protoplanetary disks.  High resolution two- and 
three dimensional simulations help us to understand the hydrodynamic 
and tidal interactions between protoplanets and disks 
\citep[e.g.,][]{kley99,lsa99,bryden99, kdh01, bate}.
However, a major shortcoming of all these codes is that they assume a 
very simple equation of state and include no radiative transfer 
effects.  
\citet{boss01} does consider radiative transfer in the diffusion 
approximation, but these are simulations of relatively massive disks 
with high accretion rates and include only compressional and viscous 
heating, whereas our models concern passive disks in which stellar 
irradiation is the primary source of heating.  
Generally speaking, 
simulated protoplanetary disks are typically vertically isothermal and  
do not include heating from the central star.  
While they 
can probe gravitational and tidal effects of a planet in a disk, 
they cannot account for effects of shadowing and illumination 
on the temperature structure as a gap opens in a disk.

Conversely, the analytic disk models self-consistently 
calculate effects of radiative 
transfer, which is important because a major source of heating in
circumstellar disks is irradiation from the central star
\citep[e.g.,][]{calvet,CG,vertstruct,DDN}.  Analytical models show 
that disk temperature structure can vary greatly with disk height 
due to heating at the surface from stellar irradiation and 
viscous heating at the midplane.  However, these models cannot account 
for perturbations in the disk such as those imposed by the 
presence of a planet because they only consider radiative transport 
in one or two dimensions.  
Monte Carlo simulations of radiative transfer on a disk with a 
gap created by a planet have been done,
but these models are essentially two-dimensional as well
\citep{rice2003,WWBW}.  These models are interesting from an 
observational point of view because large planets are able to 
significantly change the disk structure, but they do not address 
what happens to planets below the gap-opening threshold, where 
planet growth and migration are poorly understood.  

The temperature structure of a disk, particularly in the vertical 
direction, can have an important effect on the dynamics of 
disk-protoplanet interactions since waves in disks do not necessarily 
carry energy evenly with height if it is not vertically isothermal
\citep{98lubowogilvie}.  The local temperature structure 
can significantly affect how waves are dissipated in the disk, 
which will in turn affect tidal torques and migration rates.  

In this paper, we make a number of modifications to the model presented in 
\paperone{} for calculating 
radiative transfer on perturbed disks, 
primarily in the calculation of the disk properties.  
These changes are described in detail in \S\S\ref{diskstruct} and 
\ref{planet_in_disk}.  The algorithm for calculating radiative 
transfer remains essentially the same.  

In \S\ref{diskstruct}, we calculate the structure of the unperturbed 
disk, in \S\ref{planet_in_disk} we calculate the effect of a
protoplanet on the disk structure, and in \S\ref{radtrans} we 
review the method of calculating radiative transfer on a three-dimensional 
perturbation in the disk.  In \S\ref{results} 
we apply the revised method and analyze the results.  
We compare the results with those previously obtained in \paperone{}, 
calculate the effect 
on the temperature of the disk photosphere over varying planet masses 
and distances, and discuss how these temperature variations 
change the locations where ice can form in the disk.  
Section \ref{discussion} is a discussion of the results and implications.

\section{Disk Structure}

We assume that the gas and dust in the disk is well-mixed, with the 
dust primarily responsible for the opacity.  The disk is flared, 
and has a temperature inversion due to radiative heating at the 
disk surface from the central star.

\label{diskstruct}
To calculate the unperturbed disk structure, we adopt the formalism 
developed by \citet{calvet} and 
\citet{vertstruct,dalessio2}, with some simplifying 
assumptions.  The parameters we use to describe the disk structure are 
the density $\rho(r,z)$, temperature $T(r,z)$, and 
optical depth $\tau(r,z)$.  We define $\tau$ without a subscript to be 
the optical depth of the 
disk to its own radiation perpendicular to the disk.  

We assume that the disk is locally plane parallel to decouple the 
radial and vertical dependencies of the disk properties.  
For a given radius $r$, the vertical structure is calculated as follows.  
The optical depth is given by 
\begin{equation}
\tau(z) = \int_z^{z_{\infty}} \chi_R \rho(z') dz'.
\end{equation}
The density and temperature are calculated assuming 
hydrostatic equilibrium,
\begin{equation}
\frac{dP}{dz} = -\rho g_z
\end{equation}
where $P$ is the pressure and $g_z$ is the gravitational 
acceleration perpendicular to the disk.  We assume the ideal gas law, 
$P=\rho kT/\bar{m}$, where $k$ is the Boltzmann constant, $T$ is the 
temperature, and $\bar{m}$ is the mean molecular weight of the gas, 
which we assume to be primarily molecular hydrogen.  

\citet{vertstruct,dalessio2} account for turbulent and convective 
energy transport, as well as radiative transfer in their disk models; 
however, they find that radiative transfer is the primary mechanism 
for energy transport, especially in the upper layers of the disk.  
Indeed, while \citet{calvet} use a simplified set of equations 
for calculating the disk structure, their results give a close 
match to \citet{vertstruct}.  For these reasons, we shall ignore 
turbulent and convective fluxes and use the simpler set of equations 
from \citet{calvet}.

The temperature in the disk as a function of optical depth and angle 
of incidence of stellar radiation $\mu_0$ can be expressed as 
\begin{equation}
T(\tau,\mu_0) = [T_v^4(\tau) + T_r^4(\tau,\mu_0)]^{1/4}
\end{equation}
where $T_v$ and $T_r$ are temperatures due solely to viscous heating and 
stellar irradiation, respectively.

We assume that viscous flux is generated at the midplane and 
transported radiatively in a grey atmosphere so that 
\begin{equation}
T_v = \left[\frac{3F_v}{8\sigma_B}(\tau+2/3)\right]^{1/4}
\end{equation}
where $\sigma_B$ is the Stefan-Boltzmann constant.
The viscous flux $F_v$ at a distance $r$ for a star of 
mass $M_{\star}$ and radius $R_{\star}$ 
accreting at a rate $\dot{M}_a$ is 
\begin{equation}
F_v = \frac{3GM_{\star}\dot{M}_a}{4\pi r^3}
	\left[1-\left(\frac{R_{\star}}{r}\right)^{1/2}\right]
\end{equation}
\citep{pringle}.

To get $T_r$, we adopt the equations for the flux and mean intensity 
for the diffuse stellar radiation field ($F_s$ and $J_s$) 
and the differential equations for the flux and mean intensity of 
the disk's own radiation ($F_d$ and $J_d$) from \citet{vertstruct}.  
We then solve the differential equations for $F_d$ and $J_d$ using the 
boundary condition for relating disk and stellar fluxes, as in 
\citet{calvet}:
\begin{equation}
F_s + F_d = F_{\mbox{\footnotesize irr}} e^{-\tau_s/\mu_0}
\end{equation}
Solving, we obtain the same equations for the temperature as in 
\citet{calvet} with 
$q\equiv\chi_P^{\star}/\chi_R$, except
\begin{eqnarray}
C_2' &=& \frac{(1+C_1)}{\mu_0} 
	\left(\frac{\chi_P^{\star}}{\kappa_P}-\frac{3\mu_0^2}{q}\right), \\
C_3' &=& C_2 \beta 
	\left(\frac{\chi_P^{\star}}{\kappa_P} - \frac{3}{q\beta^2}\right).
\end{eqnarray}

We use the opacities from \citet{dalessio3} using a dust model with 
parameters 
$a_{\mathrm{max}} = 1\,\mbox{mm}$ and $p = 3.5$.  For simplicity, we 
assume that the dust opacities are constant throughout the disk, even 
though disk temperatures are typically above the ice sublimation point 
close to the star and at the surface of the disk, and below 
the sublimation point
farther from the star and near the midplane.  To test our assumption 
of constant opacities, we 
calculate the our models using the opacities at $T=100$
and 300 K and typically find that the results do not change 
much.  

The upper boundary condition is set so that 
$P(z_{\infty}) = 10^{-10}$ dyne, and we integrate the equations 
for $\tau$, $\rho$, and $T$ down to the midplane using some initial 
guess for $z_{\infty}$.  The other boundary condition is that we 
match the total integrated surface density 
\begin{equation}\label{surfden_int}
\Sigma = \int_{-z_{\infty}}^{z_{\infty}} \rho dz
\end{equation}
with the surface density given by a steadily accreting viscous disk
\begin{equation}\label{surfden_vis}
\Sigma = \frac{\dot{M}}{3\pi\nu} 
	\left[1-\left(\frac{R_{\star}}{r}\right)^{1/2}\right] 
\end{equation}
\citep{pringle}.
We adopt a standard Shakura-Sunyaev viscosity with $\nu = \alpha c_s H$
\citep{shaksun}.
This boundary condition is required 
to account for the effect of viscosity on the 
disk structure, since we do not treat the propagation of 
turbulent and convective fluxes.
This approximation is justified by Fig.~7 in \citet{vertstruct}, 
which shows that the integrated 
surface density profile of the detailed calculated model closely matches 
\eqref{surfden_int}.  Depending on the difference between 
\eqref{surfden_int} and \eqref{surfden_vis}, we adjust our guess for 
$z_{\infty}$ until the values for $\Sigma$ converge.

The angle of incidence, $\mu_0$, depends on slope of surface $dz_s/dr$, 
where $z_s$ is the ``surface'' of the disk, where $\tau_s/\mu_0=2/3$.  
To get a self-consistent answer for $\mu_0$ we iteratively calculate 
the vertical structure of the disk at intervals of 
$\Delta\log r = \frac{1}{2}\log 2$, calculating the slope of the surface 
between intervals of $r$.  

\begin{figure}[htbp]
\resizebox{\textwidth}{!}{\includegraphics{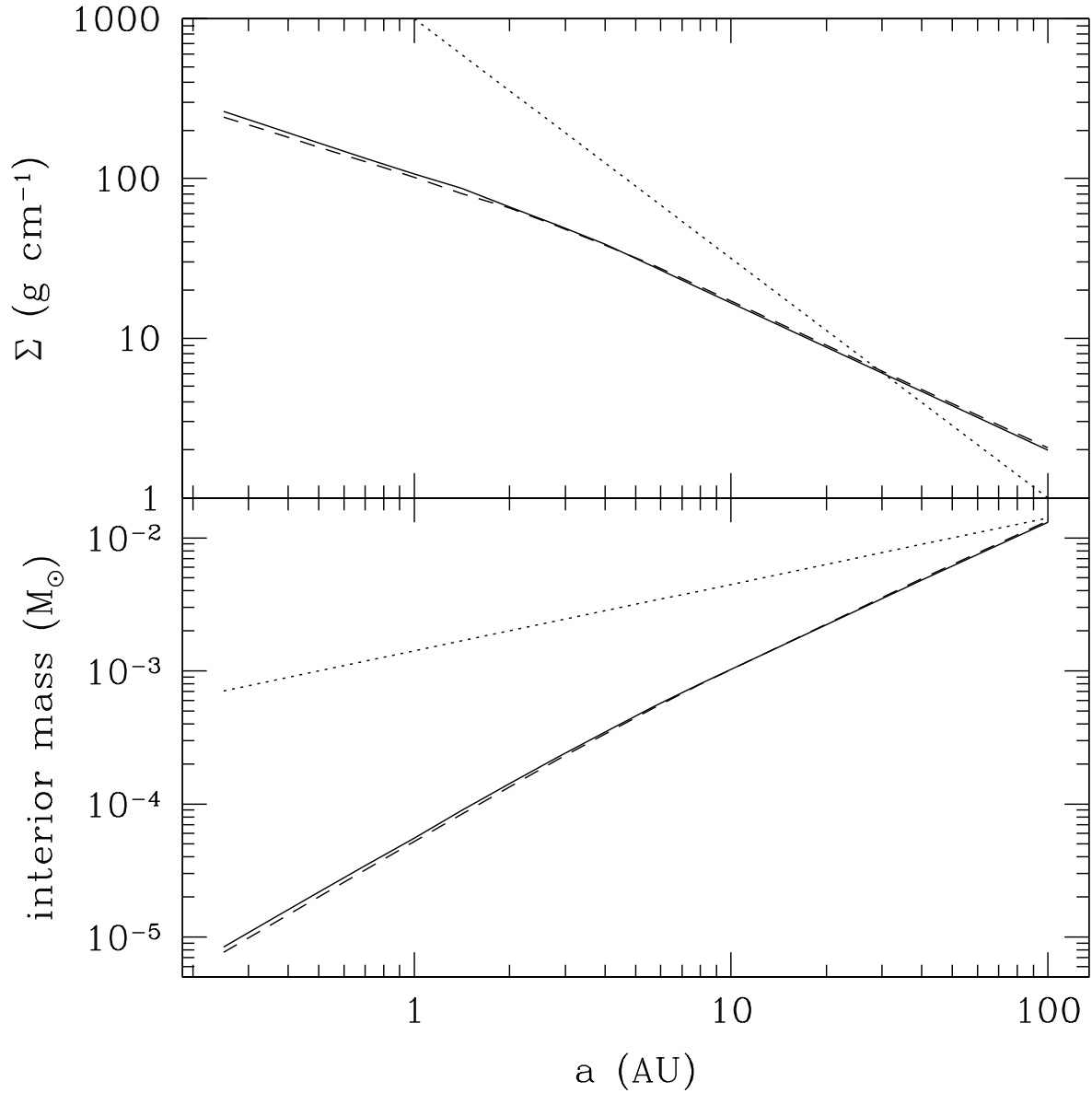}}
\caption{\label{massprof}Surface density (upper) and mass (lower)
profiles of the structure calculated for the fiducial disk model.  
The solid and dashed lines represent models using opacities for 
dust at 300 K and 100 K, respectively.  
Profiles beyond 4 AU are extrapolated.  
For comparison, the profiles for 
a disk with $\Sigma\propto r^{-3/2}$, normalized so that 
the enclosed disk mass at 50 AU is $0.01\msun$, are plotted as 
dotted lines.
}
\end{figure}

\begin{figure}[htbp]
\resizebox{\textwidth}{!}{\includegraphics{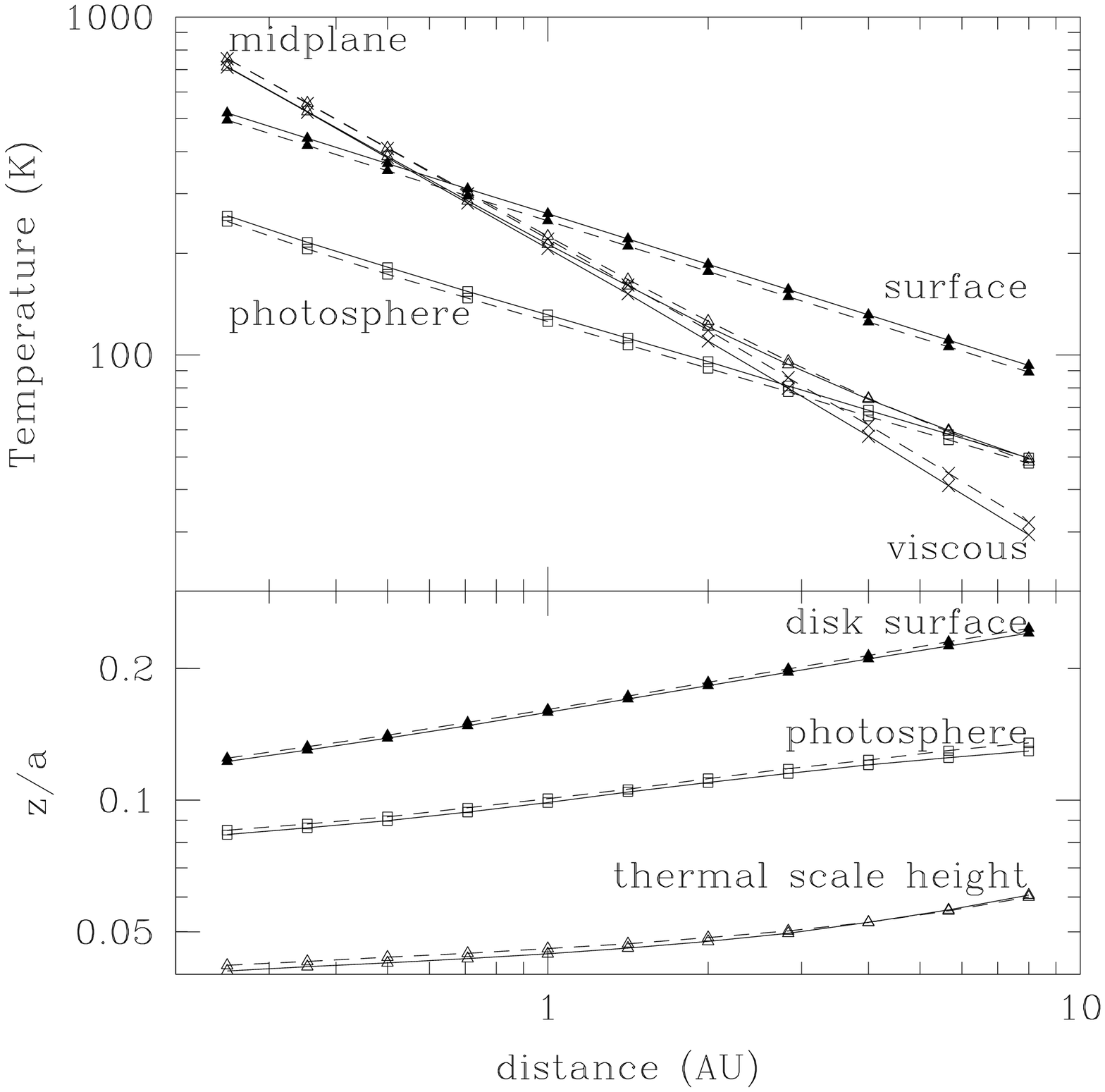}}
\caption{\label{tempht}
Temperature and vertical profiles at the indicated heights in the disk.  
The solid and dashed lines represent the 300 K 
and 100 K dust models, respectively, in both the upper and lower plots.  
{\em Upper panel}: The temperature profiles are represented by 
filled triangles for the surface, 
open squares for the photosphere, 
open triangles for the midplane, 
and crosses for the viscous temperature at the photosphere.
{\em Lower panel}: The values of various disk heights in the disk 
are represented by 
filled triangles for the surface, 
open squares for the photosphere,
and open triangles for the thermal scale height.
}
\end{figure}

For our fiducial disk-star system, we take 
$M_{\star} = 0.5\msun$, 
$R_{\star} = 2\:R_{\sun}$, 
$T_{\star} = 4000\:K$,
$\dot{M} = 10^{-8}\msun\:\mbox{yr}^{-1}$, 
and $\alpha=0.01$.
We calculate the structure of the disk from 0.25 to 4 AU.
Figures \ref{massprof} and \ref{tempht} summarize the features of this 
circumstellar disk.  
In \figref{massprof}, we plot the surface 
density and mass profile of the disk compared to a typical 
minimum mass solar nebula (MMSN) with $\Sigma\propto r^{-3/2}$ and 
enclosed disk mass of $0.01\msun$ at 50 AU.  Beyond 4 AU, we have 
extrapolated the surface density and mass profiles, assuming 
a power law.  The power-law slope is much shallower for the 
calculated disk than for the MMSN model, so we expect much more of 
the disk's mass to be at larger radii than the MMSN model would 
predict.  The two dust models are differentiated by line type: 
solid for 300 K, and dashed for 100 K.  

In \figref{tempht}, we show the temperature profile at various 
heights in the disk, and the locations of those heights.  
Again, the solid and dashed lines show results using the 300 K and 
100 K dust models, respectively.  
The surface 
of the disk is defined to be where most of the stellar radiation is 
absorbed, i.e., where $\tau_s/\mu_0 = 2/3$.  We define the photosphere 
of the disk to be where $\tau=2/3$, and the thermal scale height is 
$(c_s/v_{\phi})r$, where $c_s$ is the isothermal sound speed measured 
at the midplane and $v_{\phi}$ is the Keplerian speed.  
The upper plot shows temperatures at 
the surface ({\em filled triangles}), 
the photosphere ({\em open squares}), 
and the midplane ({\em open triangles}).  
We also plot the viscous contribution temperature at to the 
midplane temperature ({\em crosses}).  
The surface of the disk is always 
hotter than the photosphere, because the surface gets the most 
direct heating from stellar irradiation.  The photosphere is at 
many optical depths to the stellar irradiation, both due to the 
difference in opacities and because of the grazing angle of incidence 
of stellar radiation to the surface.  The midplane temperature is 
much higher than even the surface temperature at small $r$ 
because viscous heating dominates close to the star -- note 
that the midplane temperature is nearly equal to the viscous temperature 
$\lesssim1$ AU.  However, 
viscous heating falls off more rapidly than irradiation heating with 
distance, so at larger radii ($\gtrsim8$ AU), 
the midplane temperature becomes close 
to the photosphere temperature.  For optical depths $\tau\gtrsim2/3$, 
we expect $T_r$ to be nearly constant, so in the absence of viscous heating, 
the midplane and photosphere should be nearly isothermal.  

\begin{figure}[htbp]
\resizebox{\textwidth}{!}{\includegraphics{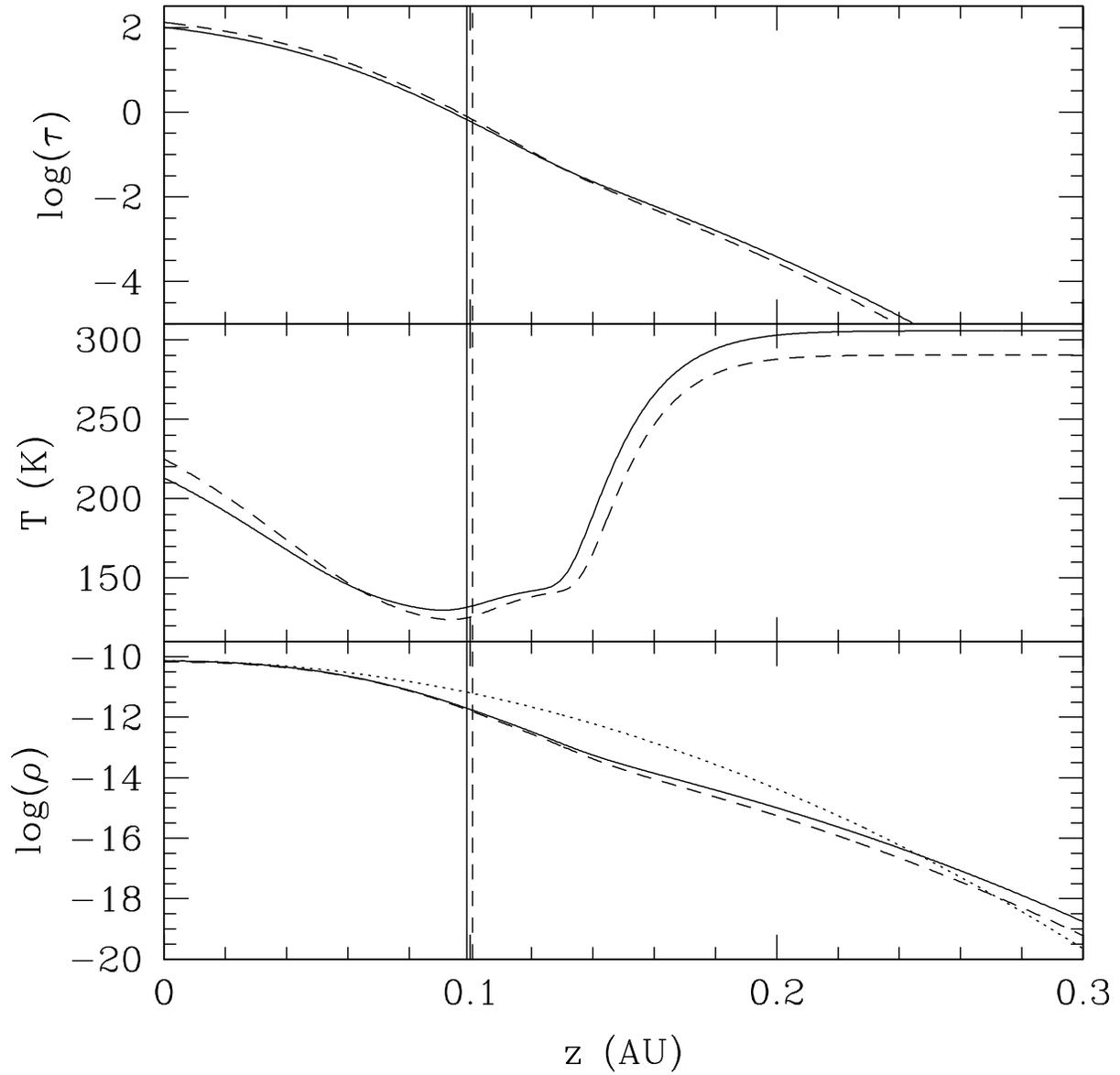}}
\caption{\label{vertprof}Vertical structure of the fiducial disk at 1 AU.
Solid and dashed lines represent 300 K and 100 K dust models, respectively.  
The vertical lines mark the locations of the photosphere.  
From top down, the plots show the variation of the optical depth,
temperature, and density of the disk vs.~disk height.
The dotted line shows the density profile of a vertically isothermal disk.
}
\end{figure}

In \figref{vertprof}, we show the vertical structure of the disk 
in optical depth, temperature, and density at a distance of 1 AU.  
As in Figures \ref{massprof} and \ref{tempht}, the solid and 
dashed lines indicate the 300 and 100 K dust models, respectively.  
The top plot shows the variation of optical depth with height 
in the disk.  The vertical lines indicate the location of the 
photosphere, where $\tau=2/3$ for the respective dust models.  
The middle plot shows the temperature structure, which rises toward the 
midplane as a result of viscous heating, reaches a minimum near the location of 
the photosphere, and rises again at the disk surface as a result of  
heating from stellar irradiation.  
The bottom plot shows the vertical density profile of the disk, 
and for comparison we have plotted the profile of a vertically isothermal
disk at 220 K as a dotted line.  
The temperature 
affects the density structure by increasing the density as temperature 
decreases, and vice versa.  
The density in our disk model 
can differ from an isothermal model by an order of magnitude 
or more due to the variations in temperature, so it is important 
to accurately and self-consistently solve for the temperature and 
density profiles.  

Overall, the difference between the 100 and 300 K dust models is very 
small, at the level of a few percent.  This justifies our assumption that 
the opacity is constant throughout the disk, and we assume 
the dust model of 300 K for the remainder of the paper.

\section{Protoplanet-Induced Density Perturbations}
\label{planet_in_disk}

We calculate the effect of a protoplanet on the disk in 
hydrostatic equilibrium in the vertical direction.  
The unperturbed density and pressure structure, $\rho$ and $P$, satisfy
\begin{equation}
\frac{1}{\rho}\frac{dP}{dz} = -\frac{G M_{\star}z}{r^3}.
\end{equation}
We express the perturbed density and temperature structure, 
$\rho'$ and $P'$, as 
\begin{equation}\label{pertHSE}
\frac{1}{\rho'}\frac{dP'}{dz} 
= -\frac{G M_{\star}z}{r^3} - \frac{G m_p z}{(x^2+y^2+z^2)^{3/2}}
\end{equation}
Now,
\[ \frac{1}{\rho}\frac{dP}{dz} = \frac{kT}{\bar{m}} \left(
	\frac{d\ln\rho}{dz} + \frac{d\ln T}{dz} \right) . \]
In general, 
\[ \left| \frac{d\ln\rho}{dz} \right| \gg \left| \frac{d\ln T}{dz} \right| \]
so we can write 
\begin{equation}
\frac{1}{\rho}\frac{dP}{dz} \approx \frac{kT}{\bar{m}}\frac{d\ln\rho}{dz},
\quad \mbox{and} \quad 
\frac{1}{\rho'}\frac{dP'}{dz} \approx \frac{kT'}{\bar{m}}\frac{d\ln\rho'}{dz}.
\end{equation}
Calculating the temperature self-consistently is not really feasible, so 
we set $T' \approx T(z_0)$, the midplane temperature,
then \eqref{pertHSE} integrates to
\begin{equation}
\rho' = C \rho \exp\left[
\frac{1}{c_s^2}\frac{G m_p}{(x^2+y^2+z^2)^{1/2}} \right]
\end{equation}
where $c_s$ is the midplane isothermal sound speed.
For simplicity, we assume that the midplane density is unchanged, so 
\begin{equation}
\rho' = \rho \exp\left\{
\frac{G m_p}{c_s^2}\left[
\frac{1}{(x^2+y^2+z^2)^{1/2}} - \frac{1}{(x^2+y^2)^{1/2}}
\right]\right\}
\end{equation}

\begin{figure}
\resizebox{\textwidth}{!}{\includegraphics{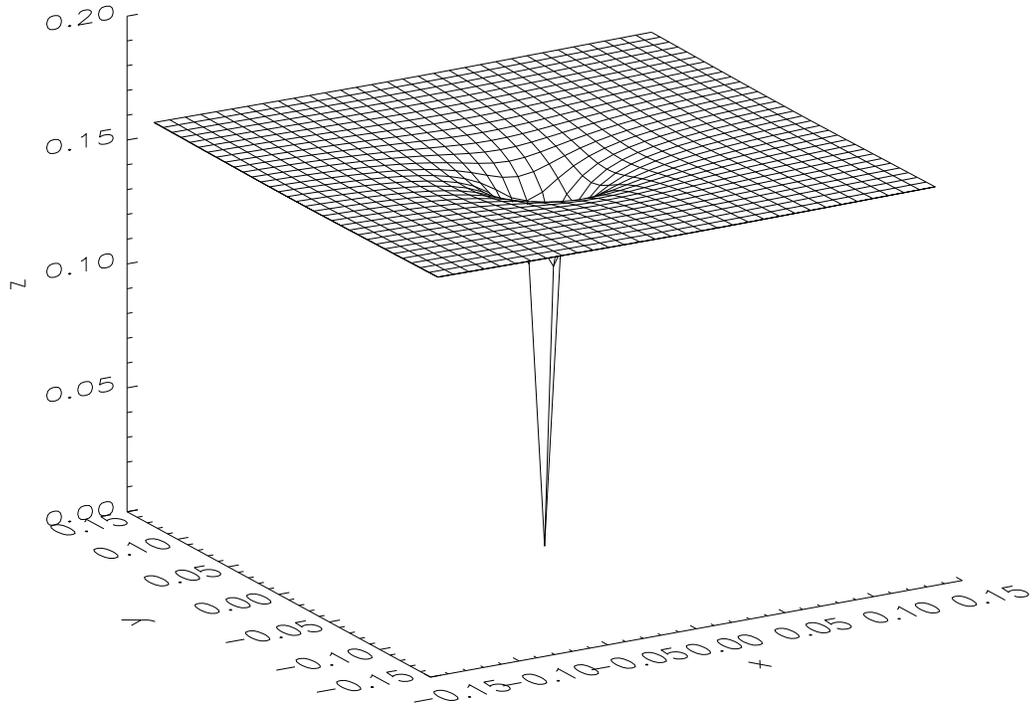}}
\caption{\label{well}
The shape of the perturbed surface for a 11 \mearth{} planet at 1 AU in 
the fiducial disk.  All units are in AU.
}
\end{figure}

If we define the isodensity contour at the density of the unperturbed 
disk surface to be the perturbed surface, the shape of the perturbation
is a depression, as shown in \figref{well}.  This example is for a 
11 \mearth{} planet at 1 AU in the fiducial disk, so that the Hill radius is 
0.63 of the thermal scale height.  When this surface is irradiated at 
grazing incidence, as from a central star, one side of the well will 
be shadowed and the other side will experience more direct illumination.  
The effect on the temperature structure due to radiative heating 
and cooling is calculated as described in the following section.

\section{Radiative Transfer on Perturbations}
\label{radtrans}
\newcommand{\Btot}{B_{\mbox{\footnotesize tot}}}

\begin{figure}[htbp]
\resizebox{\textwidth}{!}{\includegraphics{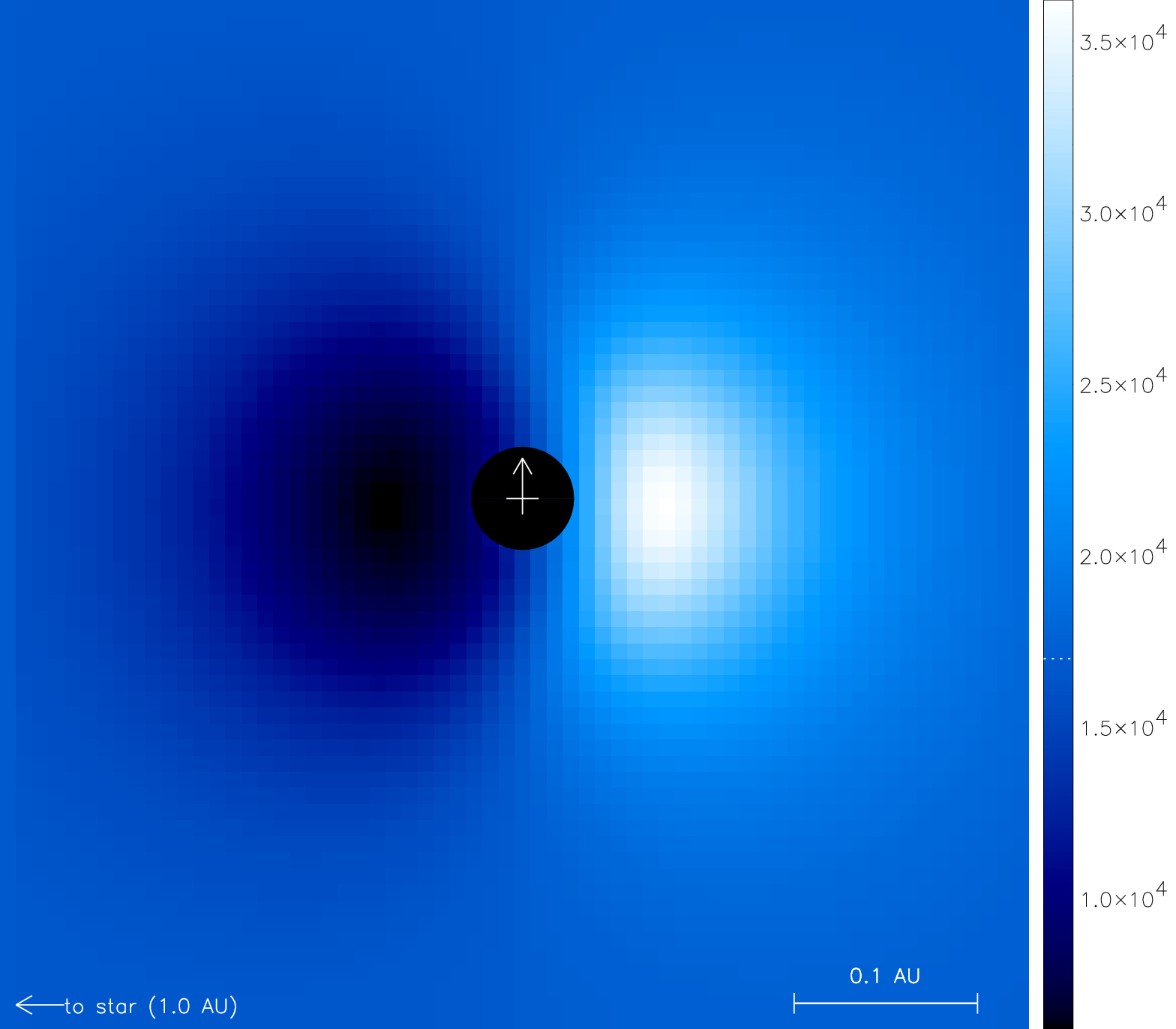}}
\caption{\label{fluxpic}The illumination pattern around a 
11 \mearth{} planet at 1 AU in the fiducial disk.  The black circle 
shows the extent of the Hill radius, with the white cross indicating 
the position of the planet.  The upward-pointing arrow shows the 
direction of the motion of the planet.  The color bar indicates 
the scaling of $\Btot{}$ in units of energy flux.
}
\end{figure}

To calculate radiative transfer on a perturbation, we use the method 
outlined in \paperone{}.  We define the surface of the perturbation 
to be the isodensity contour at the density of the unperturbed 
surface, and numerically integrate the total contributions to the 
radiative flux at a given point over this surface using the 
equation 
\begin{equation}
B_{\mathrm{tot}} = \frac{1}{\pi} \int B(\tau_d,\mu)\,\nu\;\delta\Omega 
\label{btot}
\end{equation}
We refer to the spatial variation of 
$\Btot{}$ as the illumination 
pattern in the disk.  In \figref{fluxpic}, we illustrate the illumination 
pattern in the photosphere of the fiducial disk at 1 AU in the vicinity 
of a planet of mass 11 \mearth{}.  We can easily see the effect of shadowing 
and illumination in terms of lower or higher values of $\Btot{}$. 

\begin{figure}[htbp]
\resizebox{\textwidth}{!}{\includegraphics{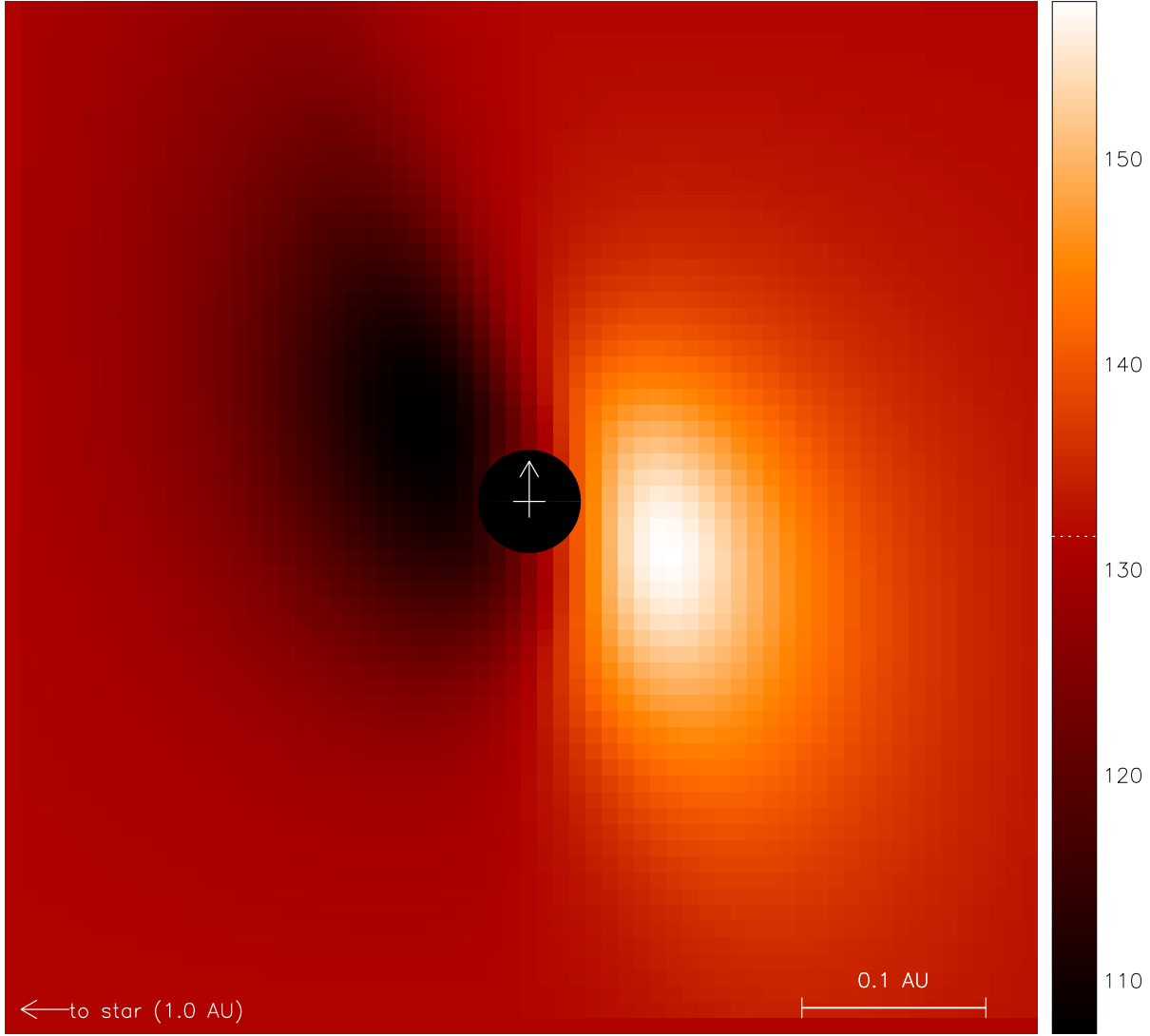}}
\caption{\label{temppic}The temperature profile around a 
11 \mearth{} planet at 1 AU in the fiducial disk.  Markings on the plot 
are the same as \figref{fluxpic}, except that the color bar here 
indicates the temperature scale in kelvins.
}
\end{figure}

To calculate the temperature distribution, we assume that the gas travels 
on streamlines in Keplerian orbits, which is an adequate assumption 
outside the Hill radius.  The gas 
heats and cools radiatively according to 
\begin{equation}
C \frac{\partial T}{\partial t} = \Btot{} - \sigma T^4
\end{equation}
where $C$ is the specific 
heat per unit surface area of the disk, defined as
$C = k\Sigma/\bar{m}$ where $\Sigma$ is the total surface density.
The resulting temperature profile for the disk-planet system 
shown in \figref{fluxpic} is shown in \figref{temppic}.

\section{Results}
\label{results}

\subsection{Comparison of Results with Paper I and Estimate of Model Uncertainties}
\label{compare}

In \S\S\ref{diskstruct} and \ref{planet_in_disk}
we discussed the changes we have made 
to the disk model of \paperone{}.  Although we now directly 
calculate the disk surface from the isodensity contour instead of 
approximating its shape, the shape of the perturbation surface
changes very little.  However, the detailed calculation of 
disk structure including more realistic opacities for circumstellar 
disks do have an important effect.  

\begin{figure}
\resizebox{\textwidth}{!}{\includegraphics{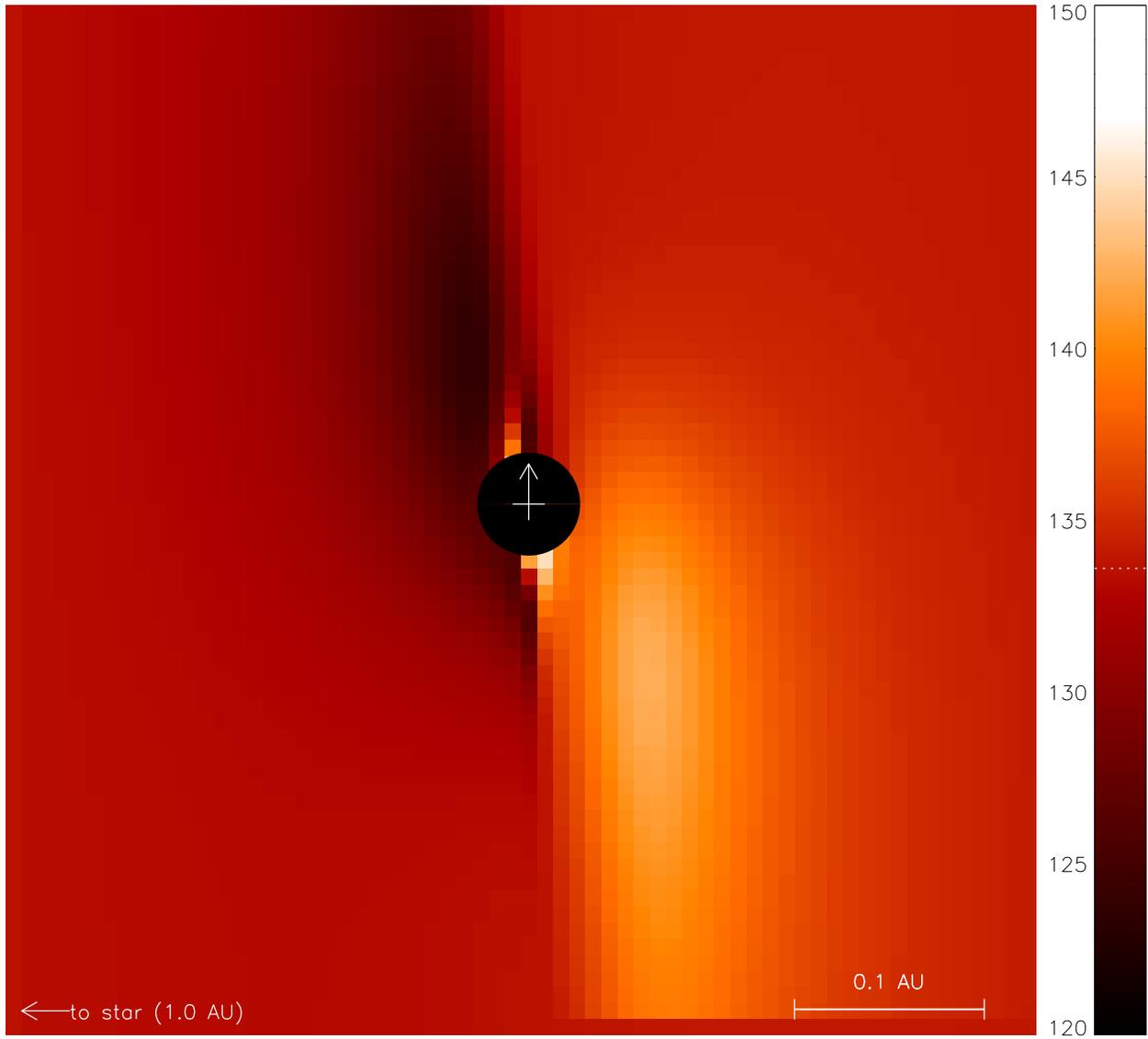}}
\caption{\label{paper1f7} Temperature profile of the disk photosphere 
near a 10 \mearth{} planet at 1 AU in orbit around a star of 
0.5 M$_{\sun}$, taken from \paperone{}, Fig.~7
}
\end{figure}

For comparison, we reproduce the temperature profile in the disk 
photosphere for a 10 \mearth{} planet at 1 AU from \paperone{} in 
\figref{paper1f7}.  Comparing this to \figref{temppic}, two differences 
are immediately obvious.  The area over which the temperature is 
perturbed is much larger in \figref{temppic}, particularly in the 
radial direction.  Also, the magnitude of the perturbation is much larger,
with temperatures varying from 107 K to 158 K in \figref{temppic},
versus 124 K to 142 K in \figref{paper1f7}.

The new method of calculating disk structure gives a much lower 
surface density at 1 AU than one would expect from a MMSN disk, 
which we had previously used to find the surface density, 
as can be seen from \figref{massprof}. 
This means a smaller specific heat, 
which means a faster radiative heating/cooling rate, which means 
less smearing out of the temperature gradient from differential 
disk rotation.  The shadow and illuminated region cool and heat 
more effectively; hence, the larger size of the temperature 
perturbation in the photosphere.  

The new opacities adopted in this paper are much 
are smaller than those used in \paperone{}, which used 
opacities representative of interstellar medium dust 
rather than disk dust.  
Since the optical depths are 
smaller, this means that the photosphere is deeper,
that is, farther away from the surface.  This increases the amount 
of solid angle that the shadowed or illuminated region subtends 
in reference to a point in the photosphere.  This also helps to 
increase the area of the temperature perturbations.  

\begin{figure}
\resizebox{\textwidth}{!}{\includegraphics{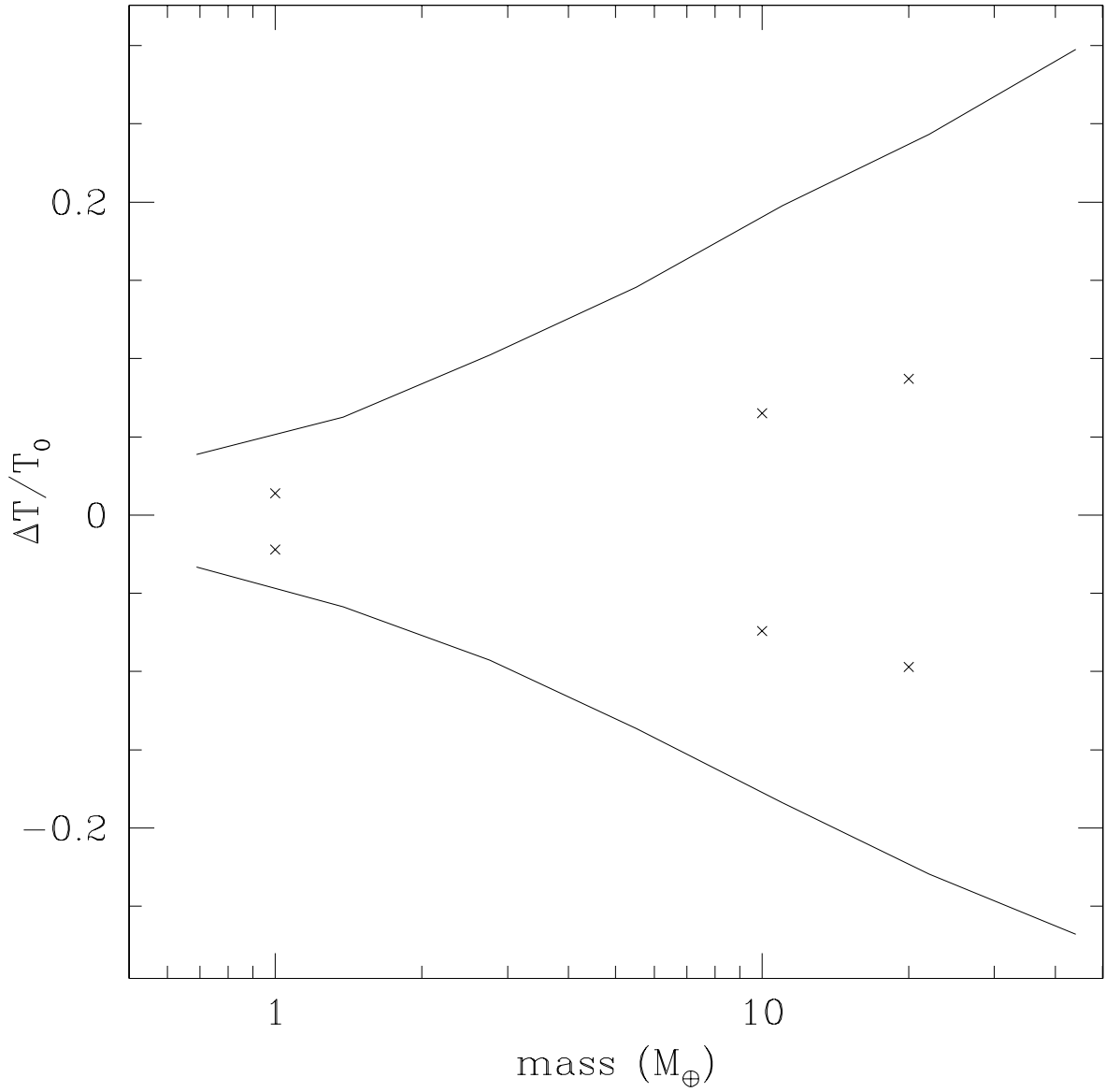}}
\caption{\label{comptemps}Comparison of temperature variations 
between \paperone{} and results presented here.  
Fractional temperature variation is plotted vs.~planet mass.
The lines show temperature maxima and minima for the parameter 
study presented in this paper, and the crosses show results 
from \paperone{}.}
\end{figure}

The new opacities also give a smaller ratio of opacities, 
which means that stellar radiation 
penetrates the disk better, increasing the effect of the irradiation 
on the temperature structure of the disk.  This is the main reason
for the increase in the magnitude of the temperature variation over 
those reported in \paperone{}, although the previously mentioned effects 
also have some effect.  
A comparison of the temperature variations is shown in 
\figref{comptemps}.  The temperature variations reported in this 
paper are $\sim2.5-3$ times greater than those reported previously.  
Also, in the previous paper, the magnitude of temperature decrements 
were consistently greater than temperature decrements.  However, this 
trend is reversed in this paper because $\mu_0$ is smaller: 
$\mu_0=0.045$ in \paperone{} and $\mu_0=0.031$ here.  The smaller 
value of $\mu_0$ means that the disk gets less direct radiation in the 
unperturbed state, so that regions that get more direct illumination
from the star get relatively hotter.

In summary, the comparison between the results of \paperone{} 
and this paper illustrate the range of model uncertainties 
involved in calculating radiative transfer in disks.  
In this paper, 
we have tried to assume parameters for our model that accurately 
represent conditions that are observed in real disks; however, 
those assumptions are subject to change as increased telescope 
resolution and sensitivity reveal more about disks.

\subsection{Results of the Parameter Study}
\label{paramstudy}

In this section, we present the results of varying the mass 
and distance of the planet within the same fiducial disk.  

The gap-opening threshold is where the Hill radius is the same as the 
thermal scale height of the disk.  For this reason, we parameterize 
the mass of a planet in terms of $\rhill/h$, 
where $h$ is thermal scale height of the disk.  We examine planets 
at distances of $0.5$, 1, 2, and 4 AU, with $\rhill/h$ ranging from 
$0.25$ to 1.  Table \ref{paramtab} summarizes the distances and masses 
of the planets examined in our parameter study.   

\begin{table}[htbp]
\caption{\label{paramtab}Planet masses, in $M_{\oplus}$}
\begin{tabular}{ccccccccc}
\tableline\tableline
$r$& \multicolumn{7}{c}{
$\rhill/h$ 
}\\
(AU) & 0.25 & 0.31 & 0.40 & 0.50 & 0.63 & 0.79 & 1.00 \\ \tableline
0.5 & 0.596 & 1.19 & 2.38 & 4.77 & 9.53 & 19.1 & 38.1 \\
1 &   0.689 & 1.38 & 2.76 & 5.51 & 11.0 & 22.0 & 44.1 \\
2 &   0.838 & 1.68 & 3.35 & 6.70 & 13.4 & 26.8 & 53.6 \\
4 &    1.13 & 2.26 & 4.51 & 9.03 & 18.1 & 36.1 & 72.2 \\
\tableline \tableline
\end{tabular}
\end{table}

\begin{figure}[htbp]
\resizebox{\textwidth}{!}{\includegraphics{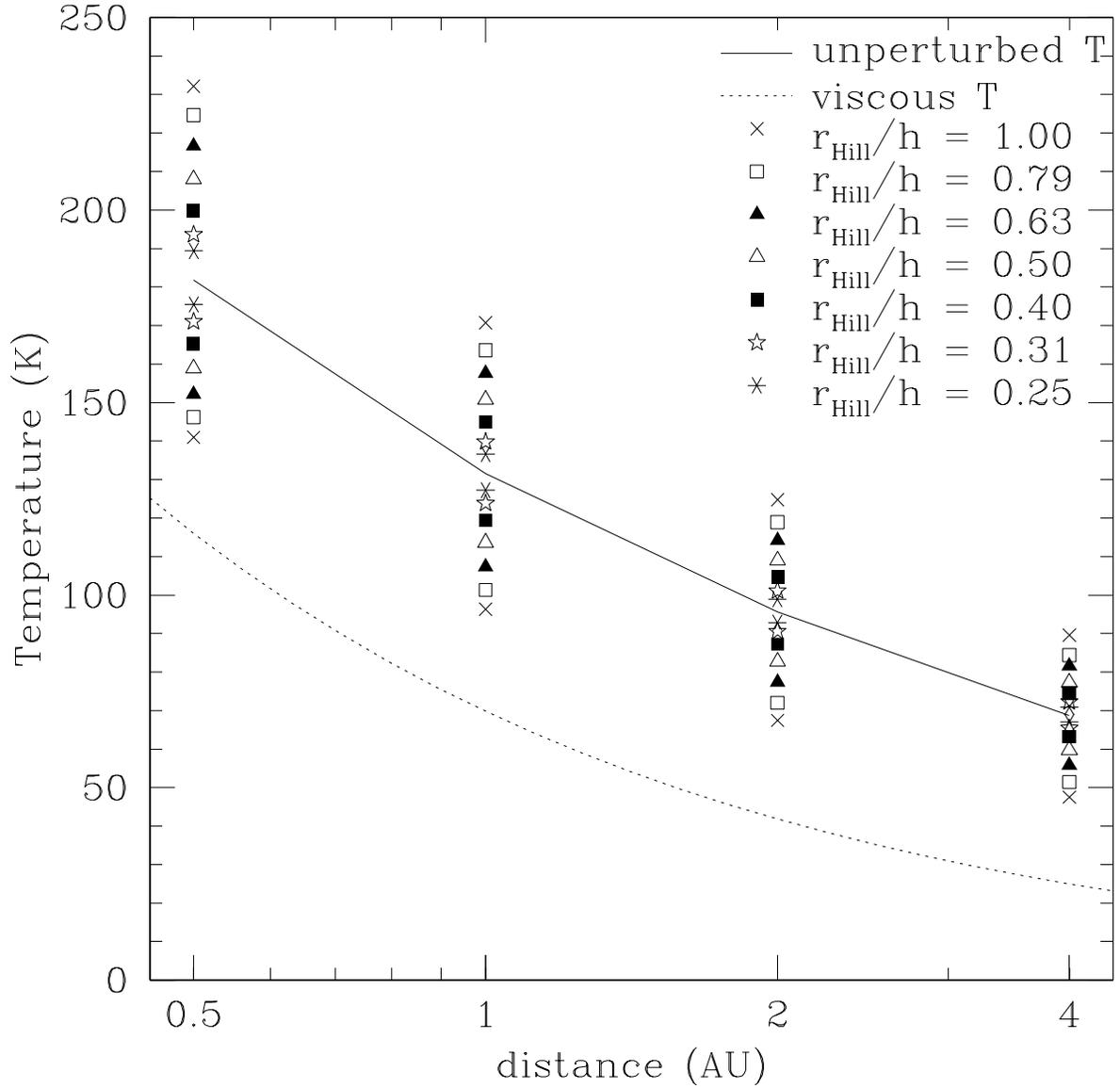}}
\caption{\label{t_vs_rad}Temperature vs.~distance.  The solid line shows 
unperturbed temperature, and the dotted line shows the viscous temperature. 
The symbols below and above the line represent minimum and maximum 
temperature, respectively, for the indicated planet size.}
\end{figure}

\figref{t_vs_rad} summarizes the variations in temperature produced by 
changing planet mass and distance.  The solid line represents the 
unperturbed photospheric temperature versus radius.  The symbols represent 
different planet masses, parametrized by the ratio of the Hill radius 
to the disk scale height.  As indicated, planets can induce temperature 
variations of several tens of K, and still be below the gap-opening 
threshold.  The minimum photospheric temperature at any given radius,
i.e., in the absence of radiative heating from the star,
is the viscous temperature at $\tau=2/3$, given by 
$T_v = (F_v/2\sigma_B)^{1/4}$.  This temperature is 
indicated by the dotted line.  
At small distances, the minimum temperatures become limited 
by viscous heating, since viscous heating contributes more to the 
photospheric temperature at smaller radii.  

\begin{figure}[htbp]
\begin{center}
\resizebox{3.2in}{!}{\includegraphics{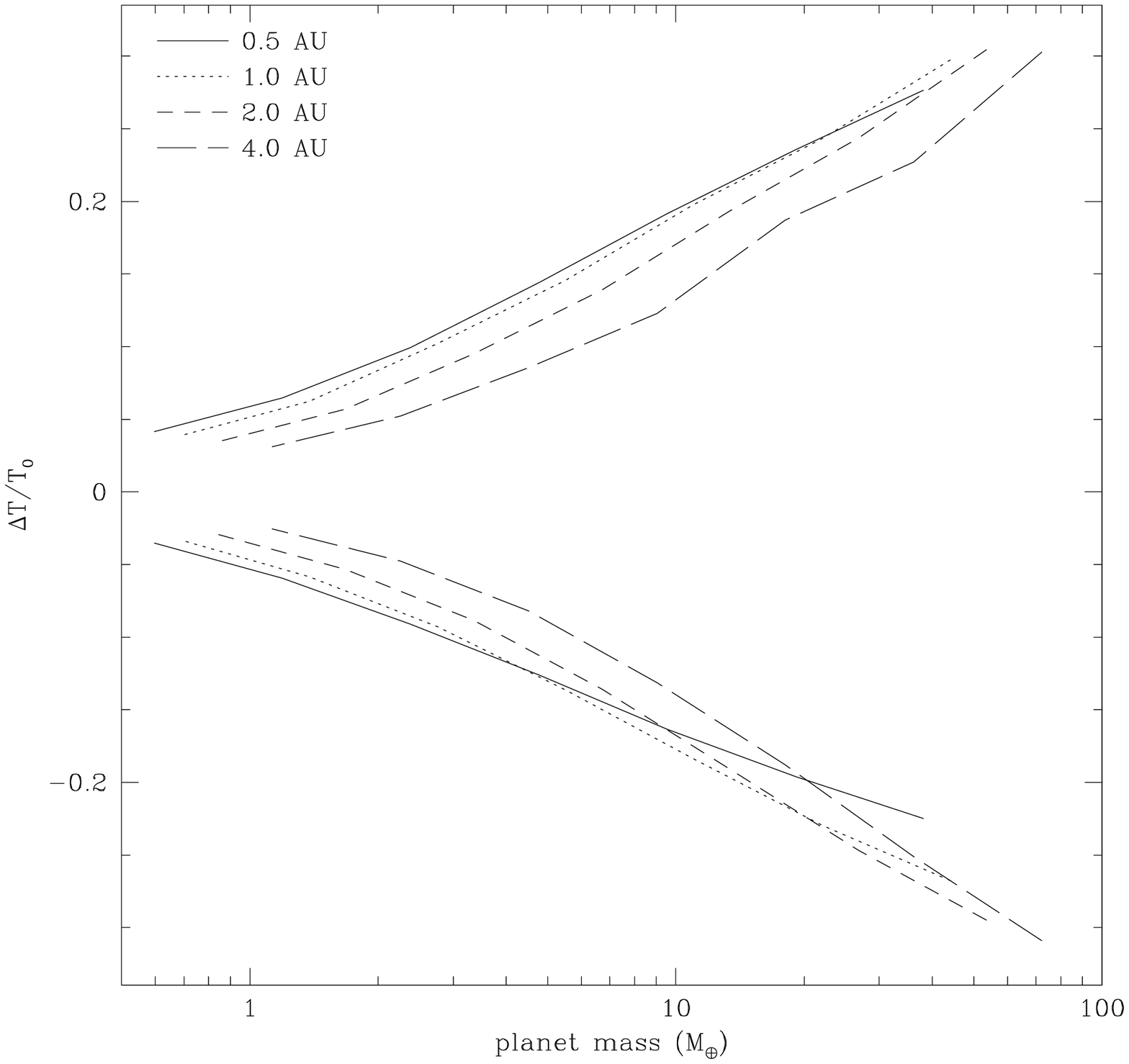}}
\resizebox{3.2in}{!}{\includegraphics{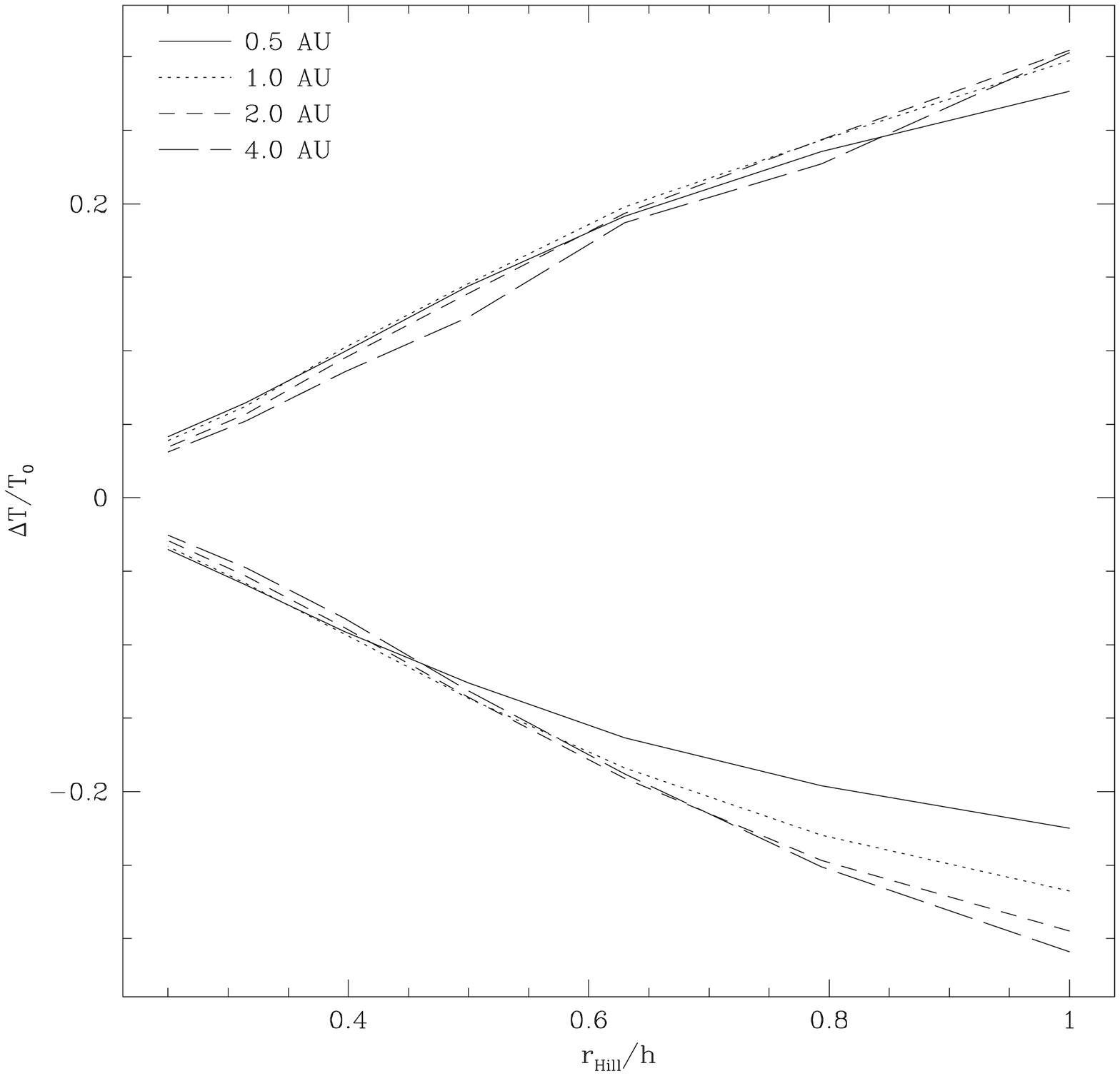}}
\end{center}
\caption{\label{t_vs_mass}Fractional temperature variation vs.~planetary 
size. {\em Left}: Horizontal axis is planet mass in \mearth{}.
{\em Right}: Horizontal axis is $\rhill/h$.}
\end{figure}

\figref{t_vs_mass} ({\em left}) 
shows the fractional temperature variation versus 
planet mass.  Each line represents a different distance to the star.  
The lines above and below $\Delta T/T_0=0$ show maximum 
and minimum temperatures, respectively.  
The temperature variation can be as much as $\pm30\%$ for the 
largest planets studied.  
If we plot the fractional temperature variation versus $\rhill/h$, 
as in \figref{t_vs_mass} ({\em right}), the lines lie almost coincident, 
indicating that the relevant scale is $\rhill/h$, not the planet mass.
The divergence of the minimum temperature toward higher $\rhill/h$ 
can be explained by the 
rise in viscous heating as the distance to the star becomes smaller.  
Since only stellar irradiation heating is changed by the presence of 
a planet in our model, viscous heating sets a lower bound to
the temperature.  The viscous temperature goes as $r^{-3/4}$ while 
the stellar radiation temperature goes as $r^{-1/2}$, so that close 
to the star, viscous heating dominates.  
 
\subsection{Hot and Cold Spots and Their Effect on the ``Snow Line''}

\begin{figure}[htbp]
\scalebox{0.75}{\includegraphics{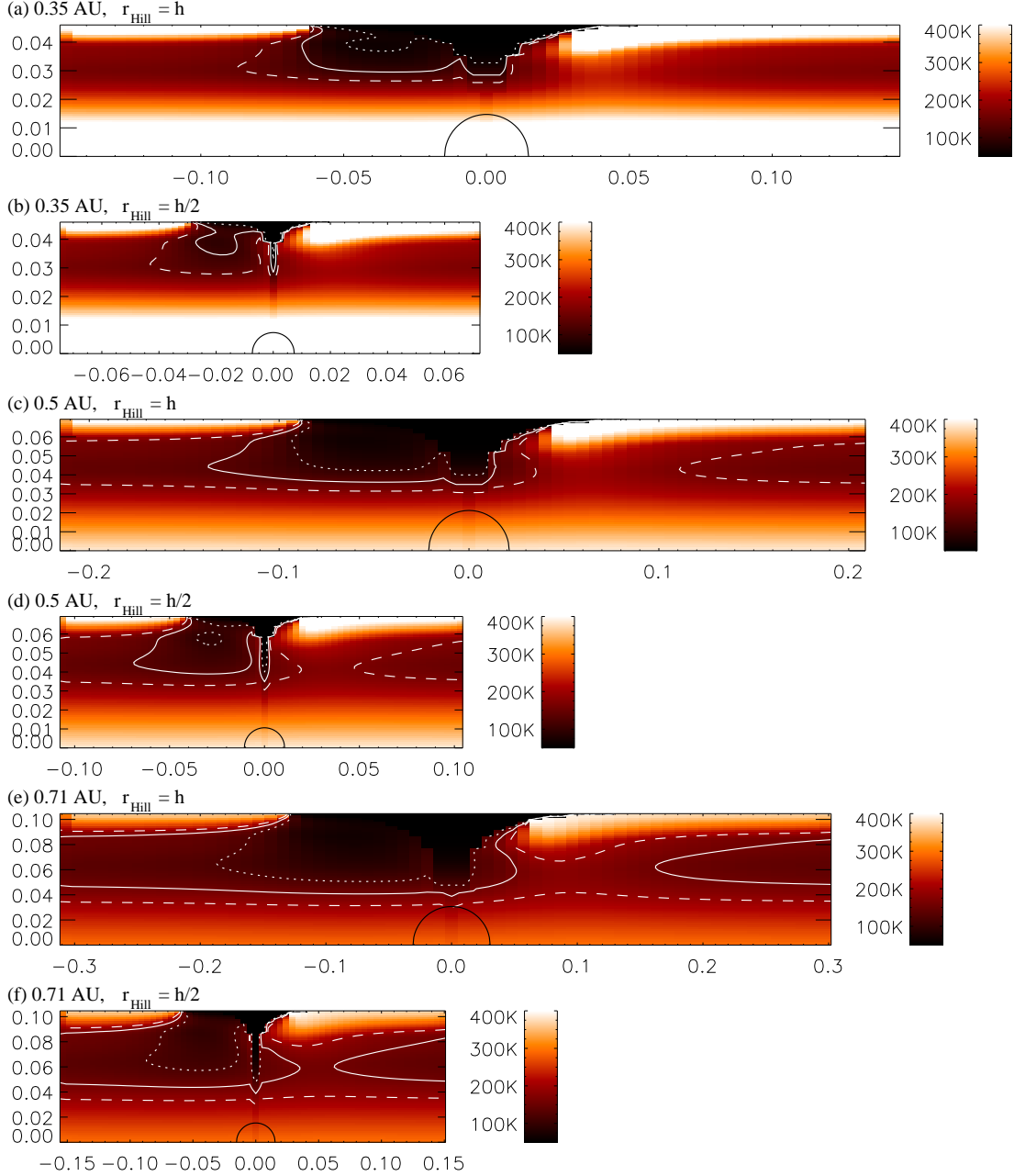}}
\caption{\label{slice1}\small Vertical temperature cross-sections for 
a selection of planet models.  The planet is located at the origin 
and the star is off to the left.  Radiation from the star 
strikes the surface at grazing incidence.  The color bars indicate 
the temperature scale.  
The white temperature contours 
correspond to 140 K ({\em dotted line}), 170 K ({\em solid line}) 
and 200 K ({\em dashed line}).  
The black semi-circle centered at the planet's position indicates the 
Hill radius.  Units on the horizontal and vertical axes are in AU, 
and indicate distance radially and vertically from the midplane, 
respectively.  
({\em a}) $m_p=35.9\:\mearth{}$ ($\rhill=h$)   at 0.35 AU, 
({\em b}) $m_p=4.49\:\mearth{}$ ($\rhill=h/2$) at 0.35 AU,
({\em c}) $m_p=38.1\:\mearth{}$ ($\rhill=h$)   at 0.5 AU, 
({\em d}) $m_p=4.77\:\mearth{}$ ($\rhill=h/2$) at 0.5 AU,
({\em e}) $m_p=40.8\:\mearth{}$ ($\rhill=h$)   at 0.71 AU, 
and 
(f) $m_p=5.1\:\mearth{}$  ($\rhill=h/2$) at 0.71 AU.}
\end{figure}

\begin{figure}[htbp]
\scalebox{0.75}{\includegraphics{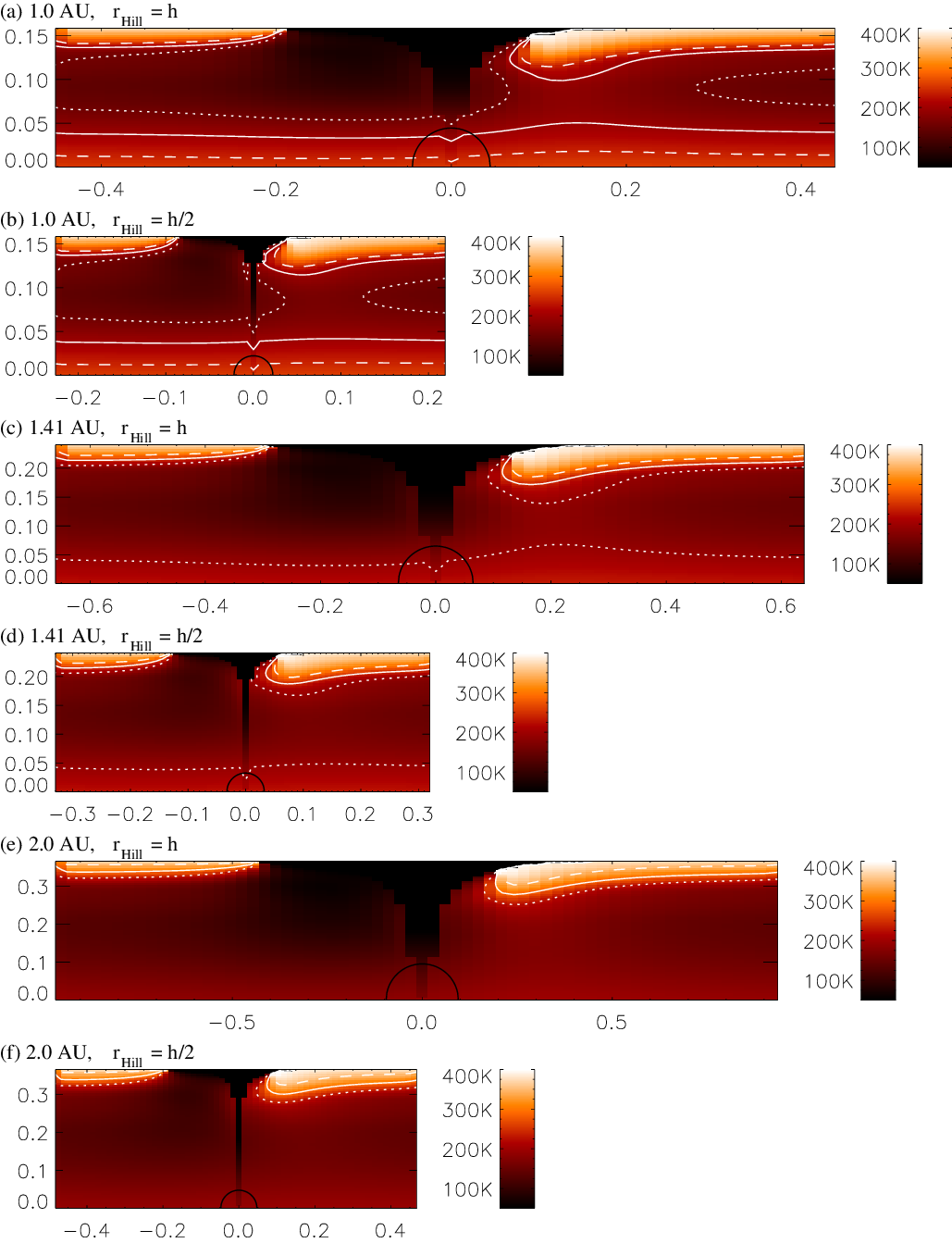}}
\caption{\label{slice2}Vertical temperature cross-sections for 
a selection of planet models.  See \figref{slice1} for details.
({\em a}) $m_p=44.1\:\mearth{}$ ($\rhill=h$)   at 1.0 AU, 
({\em b}) $m_p=5.51\:\mearth{}$ ($\rhill=h/2$) at 1.0 AU,
({\em c}) $m_p=48.5\:\mearth{}$ ($\rhill=h$)   at 1.41 AU, 
({\em d}) $m_p=6.06\:\mearth{}$ ($\rhill=h/2$) at 1.41 AU,
({\em e}) $m_p=53.6\:\mearth{}$ ($\rhill=h$)   at 2.0 AU, 
and 
({\em f}) $m_p=6.7\:\mearth{}$  ($\rhill=h/2$) at 2.0 AU}.
\end{figure}

Figures \ref{slice1} and \ref{slice2} show the vertical temperature 
cross sections for a sampling of planet masses and distances.  
At each sampled distance from the star, two plots are shown: 
a planet at the gap-opening threshold where $\rhill=h$, and a planet 
that has $\rhill=h/2$.  In each plot, 
the planet is located at the origin, with the horizontal axis in the 
radial direction and the vertical axis indicating height above the midplane. 
The units are in AU and the star is positioned to the left.  Note 
the cooling due to shadowing to the left of the planet and heating 
due to illumination on the right.
The contours indicate isotherms 
of 140 K ({\em dotted line}), 170 K ({\em solid line}) 
and 200 K ({\em dashed line}).  The contour 
corresponding to 170 K is of particular interest, because this is the 
sublimation temperature of water.  

\begin{figure}[htbp]
\resizebox{\textwidth}{!}{\includegraphics{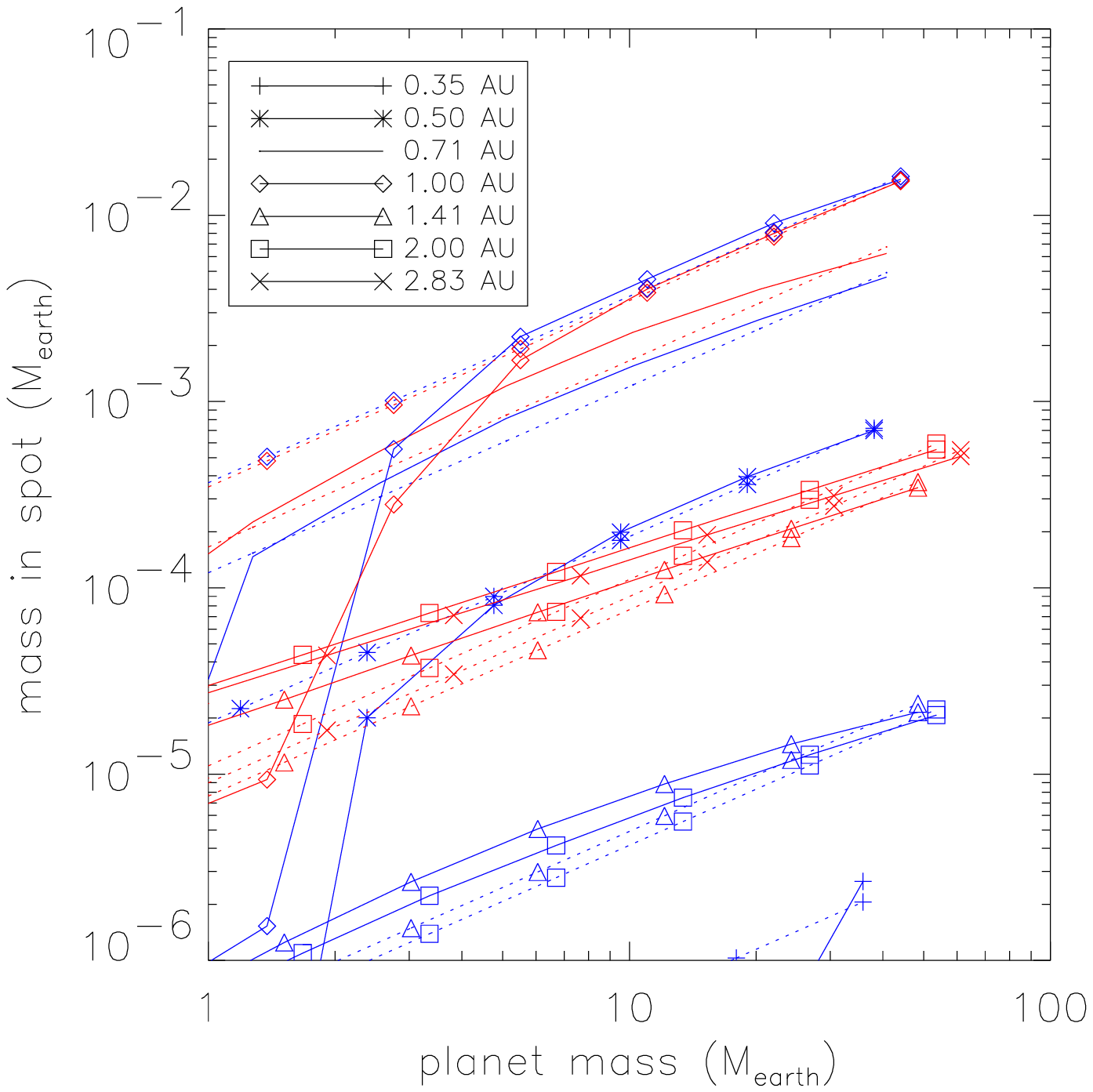}}
\caption{\label{spot_vs_planet}Masses of hot and cold spots vs.~planet 
mass.  The symbols represent different distances from the star, 
as indicated in the legend, red for hot spots and blue for cold spots.  
The solid lines represent the data points, and dotted lines represent 
the corresponding linear fits to the data.
}
\end{figure}

In this section, we shall define ``hot'' to be above 170 K, and 
``cold'' to be below 170 K.  
We define a cold (hot) spot as a region 
that is colder (hotter) than 
170 K in a layer that would be above (below) 170 K in the absence 
of a planet, and we determine the masses of the hot and cold spots 
in the disk as a function of planet mass and distance.
We find that the correlation between spot mass and planet mass is 
approximately linear, as shown in \figref{spot_vs_planet}.  The points 
connected by solid lines show spot mass versus planet mass 
at the distances indicated, blue lines for cold spots, and 
red lines for hot spots.  The corresponding
dotted lines show the linear fits 
for each of the curves.  

\begin{figure}[htbp]
\resizebox{\textwidth}{!}{\includegraphics{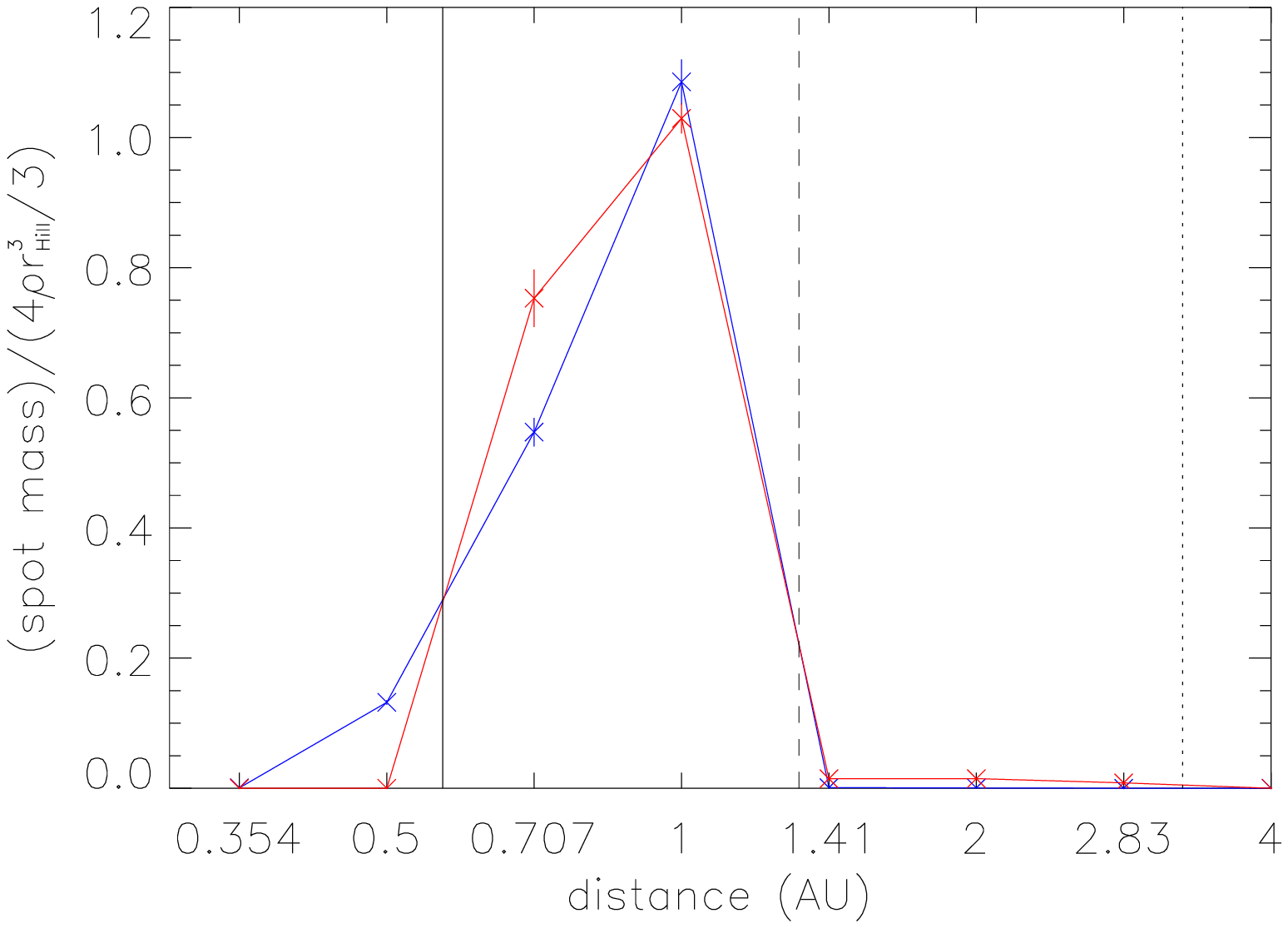}}
\caption{\label{spotratio}
Ratio of spot mass to mass of accretion zone vs.~distance.
The blue line represents the cold spot mass ratio, and the red line 
represents the hot spot mass ratio.  Errors bars are derived from 
curve-fitting.  The solid vertical line represents the beginning of the 
snow transition, where the disk begins to drop below 170 K.  
The dashed (dotted) line shows where the midplane (surface) 
temperature reaches 170 K.
}
\end{figure}

We compare the masses of the hot and cold spots to the amount of 
mass in the planet's accretion zone, which we will approximate as 
$4\rho\rhill^3/3$, the amount of disk material within the 
planet's Hill sphere.  Then the mass of the accretion zone 
also scales linearly with planet mass, since 
$4\rho\rhill^3/3 = m_p ( 4\rho a^3/9 M_{\star} )$, 
so the ratio of spot mass to the mass of the accretion zone will 
also be approximately constant.  Figure \ref{spotratio} shows 
how this ratio varies with distance from the star, for both 
hot and cold spots.  

At distances close to the star, the entire vertical extent of the disk 
will be hot.  Similarly at large distances, the entire 
vertical extent of the disk will be cold.  However, since the 
disk is not vertically isothermal, there is not a unique radius at which 
the disk is equal to 170 K, but rather there is a transition region 
where parts of the disk will be cold and the rest will be 
hot.  Hence, the ``snow line'' is not really a line, but 
rather a ``snow transition'' region as different layers in the disk 
drop below 170 K at different radii \citep{snowline}.  
Since the heating sources are at the midplane and surface, 
an intermediate layer will reach 170 K first, growing in thickness with 
increasing distance from the star, until the entirety of the disk 
is cold.  This transition is illustrated in Figures \ref{slice1}
and \ref{slice2}, where the temperature contours in the unperturbed 
parts of the plots show the extent of the snow layer.  The cold snow 
layer becomes ever thicker with increasing radius, until only the 
surface layer is hot.  
For the adopted disk model parameters, the snow transition begins 
at 0.570 AU.  The midplane temperature drops to 170 K at 1.32 AU,
and the surface reaches this temperature at 3.25 AU.

Interior to the transition region, there are no heating effects 
due to illumination of the perturbation.  However, at 0.5 AU 
there does begin to be noticeable cooling from shadowing, as shown in 
\figref{spotratio}.  This is interior to the unperturbed snow transition 
radius at 0.57 AU, indicating that the shadowing has the effect of 
moving the transition radius inward.  

The size of both hot and cold spots drops dramatically at 1.4 AU, 
roughly coincident with where the midplane temperature drops below 170 K.  
The explanation for this is that as the 170 K isotherm drops to 
lower heights in the disk, the density increases rapidly.  
So although the absolute temperature perturbation 
decreases as the optical depth increases, the relative 
amount of mass affected actually increases until about 1 AU in the 
disk.  After the midplane temperature drops below 170 K, 
radiative transfer effects 
are no longer effective enough to heat the interior 
temperature of the disk above 170 K.  Although the surface 
layer remains hot and experiences significant temperature changes 
from shadowing and illumination, the density is very low 
so the relative masses that are affected remain small.  

Since water sublimates at 170 K, the locations and masses of the hot 
and cold spots in relation to the hot and cold layers of the disk 
can have consequences for accretion of disk material onto the planet.  
The formation of ice enhances the abundance of particulate matter, which 
preferentially settles to the midplane where it can be more easily 
accreted onto the planet.  By the same token, 
hot regions will have an opposite effect. 
Thus, the existence of cold and hot regions around 
a protoplanet can enhance the rate of planetary growth and change 
the composition of disk material that is accreted.  The locations 
of the hot and cold spots are also important, since it is likely 
that planets accrete disk material assymetrically.

\section{Summary and Discussion}
\label{discussion}

We have improved on the disk model from our previous paper 
\citep{paper1} by updating the opacities and calculating the 
vertical temperature structure more self-consistently.  As a result,
temperature perturbations in the disk's photosphere 
due to the influence of a protoplanet are greater in 
magnitude.  For planets at the gap opening threshold, 
temperature variations can be up to $\pm30\%$. 

While these temperature perturbations are unlikely to be 
observed with even the most sensitive instruments, they may have 
significant effects on planet building.  The temperature variations 
are large enough to affect the composition and dynamics of the disk 
material near the planet, which can have consequences for planetary 
accretion and migration.  

If the temperature changes enough to drop above or below 
the condensation or sublimation temperature for ice formation, 
this will change the size distribution of dust grains.  This will also 
change the composition of the gas as molecules freeze out onto the dust.  
The disk temperature can also change the time scale for the dust 
settling to the midplane, so that the dust-to-gas ratio may vary 
with height.  In particular, 
shadowing and illumination effects can change 
the locations in the disk where water 
ice can form.  We have shown that ice can form inward of 0.5 AU in 
the presence of a protoplanet, whereas without the protoplanet 
the minimum distance at which ice can form is at 0.57 AU.  
This may mean that accretion rates can be enhanced closer to the 
star than previously expected.  

The temperature may also affect the movement of disk material 
near the planet by shifting the streamlines along which gas and dust 
move past the planet.  This is because of the pressure gradient that 
the temperature perturbation imposes on the disk.  Thus, accretion 
onto the planet may preferentially come from one side of the planet or 
the other.  This, along with the change in composition of disk material, 
can affect the growth rate and eventual composition of the planet.  

In addition, the temperature perturbation may change the 
migration rate of the planet under Type I migration.  
\citet{wardA} has demonstrated that the pressure gradient caused by 
typical disk temperature profiles contributes to increasing (decreasing) 
torques from the outer (inner) disk so that the total net torque 
will almost certainly cause inward migration of the planet.  
Therefore changing the temperature profile in the vicinity of the planet 
can change migrations rates.  

In future work, we will address the questions raised about planet 
growth and migration in the light of the temperature variations that we 
have studied in this paper.  

\bibliographystyle{apj}
\bibliography{apj-jour,../planets}

\end{document}